\documentclass[12pt]{article}
\usepackage{amsthm}
\usepackage{amsmath}
\usepackage{amsfonts}
\usepackage{amssymb}
\usepackage{amscd}
\usepackage{url}
\usepackage{epsfig}
\usepackage{lscape}
\usepackage{url}
\date{}
\newtheorem{thm}{Theorem}[section]
\newtheorem{cor}[thm]{Corollary}

\newtheorem{prop}[thm]{Proposition}

\theoremstyle{definition}
\newtheorem{example}[thm]{Example}
\newtheorem{remark}[thm]{Remark}
\newtheorem{lemma}[thm]{Lemma}
\newtheorem{algorithm}[thm]{Algorithm}

\newcommand{\R}{\mathbb{R}}

\newcommand{\Z}{\mathbb{Z}}
\newcommand{\N}{\mathbb{N}}
\newcommand{\Q}{\mathbb{Q}}
\newcommand{\PP}{\mathbb{P}}
\newcommand{\LL}{\mathbb{L}}
\newcommand{\B}{\mathbb{B}}

\begin{document}
\bibliographystyle{plain}
\title{Marginal Likelihood Integrals \\ for Mixtures of Independence Models}
\author{Shaowei Lin, Bernd Sturmfels and Zhiqiang Xu}
\maketitle

\begin{abstract}
Inference in Bayesian statistics involves the
evaluation of marginal likelihood integrals.
We present algebraic algorithms for computing
such integrals exactly for discrete data of small sample size.
Our methods apply to
both  uniform priors and Dirichlet priors.
The underlying statistical models are
mixtures of independent distributions, or,
in geometric language,
secant varieties of Segre-Veronese varieties. \medskip 
\end{abstract}

\section{Introduction}
\label{introduction}
Evaluation of marginal likelihood integrals is central to Bayesian statistics.
It is generally assumed that these integrals cannot be evaluated exactly, except in trivial
cases, and a wide range of numerical techniques (e.g.~MCMC) have been
developed to obtain asymptotics and numerical approximations \cite{CH}. The aim of this
paper is to show that exact integration is more feasible than is surmised in the literature.
We examine marginal likelihood integrals for a class of mixture models
for discrete data. Bayesian inference for these models arises in many contexts,
including machine learning and computational biology. Recent work in these
fields has made a connection to singularities in algebraic geometry
\cite{Drton, GR, W, YW1, YW2}. Our study augments these
developments by providing tools for symbolic integration when the sample size is small.

The numerical value of the integral we have in mind is a rational number,
and exact evaluation means computing that rational number rather than a
floating point approximation. For a first example consider the integral
\begin{equation}\label{swissintegral}
 \int_\Theta \prod_{i,j\in \{{\tt A}, {\tt C}, {\tt G}, {\tt T}\} }
 \bigl(\,\pi \lambda_i^{(1)}\lambda_j^{(2)}+\tau
\rho_i^{(1)}\rho_j^{(2)} \, \bigr)^{U_{ij}}d\pi\,d\tau\,d\lambda\, d\rho,
\end{equation}
where $\Theta$ is the $13$-dimensional polytope $\Delta_1\times
\Delta_3\times \Delta_3\times \Delta_3\times \Delta_3$.
 The
factors are probability simplices, i.e.
$$\begin{array}{rclll}
\Delta_1 &=&
\{(\pi,\tau) \in \R_{\geq 0}^2 : \pi+\tau = 1 \}, && \\
\Delta_3&=&\{(\lambda_{\tt
A}^{(k)}, \lambda_{\tt C}^{(k)}, \lambda_{\tt G}^{(k)},
\lambda_{\tt T}^{(k)})\in \R_{\geq 0}^4&: \textstyle \sum_i\lambda_{i}^{(k)}
=1 \}, &k=1,2, \\
\Delta_3&=&\{(\rho_{\tt
A}^{(k)}, \rho_{\tt C}^{(k)}, \rho_{\tt G}^{(k)},
\rho_{\tt T}^{(k)})\in \R_{\geq 0}^4&: \textstyle \sum_i\rho_{i}^{(k)}
=1 \}, &k=1,2.
\end{array}$$ 
and we integrate with respect to Lebesgue
probability measure on $\Theta$. If we take the exponents $U_{ij}$
to be the entries of the particular contingency table
\begin{equation}
\label{swisstable}
U=\begin{pmatrix}
  4 & 2 & 2 & 2 \\
  2 & 4 & 2 & 2 \\
  2 & 2 & 4 & 2 \\
  2 & 2 & 2 & 4
\end{pmatrix},
\end{equation}
then the exact value of the integral (\ref{swissintegral}) is the rational number
\begin{equation}
\label{swissnumber}
\frac{571 \cdot  773426813 \cdot 17682039596993 \cdot 625015426432626533 }{
2^{31} \cdot 3^{20} \cdot 5^{12} \cdot 7^{11} \cdot 11^8  \cdot 13^7
\cdot 17^5 \cdot 19^5 \cdot 23^5 \cdot 29^3 \cdot 31^3 \cdot 37^3 \cdot 41^3 \cdot 43^2}.
\end{equation}
The particular table (\ref{swisstable}) is taken from \cite[Example 1.3]{ASCB},
where the integrand
\begin{equation}
\label{swissintegrand}
\prod_{i,j\in \{{\tt A}, {\tt C}, {\tt G}, {\tt T}\} } \!\!\!
 \bigl(\,\pi \lambda_i^{(1)}\lambda_j^{(2)}+\tau
\rho_i^{(1)}\rho_j^{(2)} \, \bigr)^{U_{ij}}
\end{equation}
 was studied using the EM algorithm,
and the problem of validating its global maximum over $\Theta$ was raised.
 See \cite[\S 4.2]{FHRZ} and \cite[\S 3]{OpenProblems} for further discussions.
That  optimization problem, which was widely known as the
{\em $100$ Swiss Francs problem}, has in the meantime been solved
by Gao, Jiang and Zhu \cite{GJZ}.

The main difficulty in performing
computations such as (\ref{swissintegral}) = (\ref{swissnumber})
lies in the fact that the expansion of the integrand
has many terms. A first naive upper bound on the number of
monomials in the expansion of (\ref{swissintegrand}) would be
\begin{equation*}
\prod_{i,j \in \{{\tt A}, {\tt C}, {\tt G}, {\tt T}\} }  \!\! (U_{ij}+1) \quad = \quad
3^{12} \cdot 5^4 \,\,\, = \,\,\, 332,150,625 .
\end{equation*}
However, the true number of monomials
is only $ 3,892,097 $, and we obtain the rational number (\ref{swissnumber})
by summing the values of the corresponding integrals
\begin{eqnarray*}
&\displaystyle \int_{\Theta}   \pi^{a_1} \tau^{a_2}
(\lambda^{(1)})^u
(\lambda^{(2)})^v
(\rho^{(1)})^w
(\rho^{(2)})^x
d\pi\,d\tau\,d\lambda\, d\rho  \quad = &\\
&\displaystyle \frac{a_1 ! \, a_2 !}{(a_1 {+} a_2 {+} 1) !} \cdot
\frac{3!\,\prod_i u_i !}{(\sum_i u_i + 3) !} \cdot
\frac{3!\,\prod_i v_i !}{(\sum_i v_i + 3) !} \cdot
\frac{3!\,\prod_i w_i !}{(\sum_i w_i + 3) !} \cdot
\frac{3!\,\prod_i x_i !}{(\sum_i x_i + 3) !}. &
\end{eqnarray*}
The geometric idea behind our approach is that the Newton polytope
of (\ref{swissintegrand}) is a {\em zonotope} and we are summing
over the lattice points in that zonotope.
Definitions for these geometric objects are given in Section 3.

This paper  is organized as follows.
In Section 2 we describe the class of algebraic statistical
models to which our method applies, and we specify the problem.
In Section 3 we examine the Newton zonotopes of mixture models,
and we derive formulas for marginal likelihood evaluation using
tools from geometric combinatorics.
Our algorithms and their implementations are described in detail in Section 4.
Section 5 is concerned with applications in Bayesian statistics.
We show how {\em Dirichlet priors} can be incorporated into our approach,
 we discuss the evaluation of {\em Bayes factors},
 we compare our setup with that of Chickering and Heckerman in~\cite{CH},
 and we illustrate the scope of our methods by computing an integral arising
from a data set in \cite{EGG}.

A preliminary draft version of the present article was published as
Section 5.2 of the Oberwolfach lecture notes \cite{LoAS}.
We refer to that volume for further information on the use of
computational algebra in Bayesian statistics. 

\section{Independence Models and their Mixtures}

We consider a collection of discrete random variables
$$
\begin{array}{cccc}
X_1^{(1)},& X_2^{(1)},& \ldots,& X_{s_1}^{(1)}, \\
X_1^{(2)},& X_2^{(2)},& \ldots,& X_{s_2}^{(2)}, \\
\vdots & \vdots & \ddots & \vdots \\
X_1^{(k)},& X_2^{(k)},& \ldots,& X_{s_k}^{(k)},
\end{array}
$$
where $X_1^{(i)}, \ldots, X_{s_i}^{(i)} $
are identically distributed with values in  $\{0,1,\ldots, t_i\}$.
The independence model $\mathcal{M}$ for these variables is a
toric model \cite[\S 1.2]{ASCB} represented
by an integer  $d \times n$-matrix $A$ with
\begin{equation}
\label{DEFdn}
 d \, =\, t_1 + t_2 + \cdots + t_k + k \quad \hbox{and} \quad
n \, = \, \prod_{i=1}^k (t_i+1)^{s_i} .
\end{equation}
The columns of the matrix $A$ are indexed by
elements $v$ of the state space
\begin{equation}
\label{statespace}
\{0,1,\ldots,t_1\}^{s_1} \times
\{0,1,\ldots,t_2\}^{s_2} \times \cdots \times
\{0,1,\ldots,t_k\}^{s_k} .
\end{equation}
The rows of the matrix $A$ are indexed by the
model parameters, which are the $d$ coordinates of the points
$\,\theta = (\theta^{(1)},\theta^{(2)}, \ldots, \theta^{(k)})  \,$ in
the polytope
\begin{equation}
\label{prodofsimp}  P \,\,\, = \,\,\, \Delta_{t_1} \times \Delta_{t_2} \times \cdots
\times \Delta_{t_k},
\end{equation}
and the model $\mathcal{M}$ is the subset of the simplex $\Delta_{n-1}$ given
parametrically  by
\begin{equation}
\label{parametrization}
 p_v \quad = \quad
{\rm Prob}\bigl(\, X^{(i)}_j = v^{(i)}_j \,\,\hbox{for all} \,\,i,j \,\bigr) \quad =
\quad \prod_{i=1}^k \prod_{j=1}^{s_i}
\theta^{(i)}_{v_{j}^{(i)}}.
\end{equation}
This is a monomial in $d$ unknowns.
The matrix $A$ is defined by taking its column $a_v$ to be the exponent vector of this monomial.

In algebraic geometry, the model $\mathcal{M}$
is known as {\em Segre-Veronese variety}
\begin{equation}
\label{segreveronese}
\PP^{t_1} \times \PP^{t_2} \times \cdots \times \PP^{t_k}
\quad \hookrightarrow \quad \PP^{n-1},
\end{equation}
where the embedding is given  by the line bundle $\mathcal{O}(s_1,s_2,\ldots,s_k)$.
The manifold $\mathcal{M}$ is the toric variety of the polytope $P$. Both objects
have dimension $d-k$, and they are identified with each other via the moment map \cite[\S 4]{Ful}.

\begin{example}
Consider three binary random variables
where the last two random variables are identically distributed.\!
In our notation, this corresponds to
$k=2$, $s_1 = 1$, $s_2=2$ and $t_1=t_2=1$. We find that
$d = 4, n = 8$, and
$$ A \quad = \quad
\bordermatrix{ & p_{000} & p_{001} & p_{010} & p_{011} & p_{100} & p_{101} & p_{110} & p_{111} \cr
\theta^{(1)}_0 & 1 & 1 &  1 & 1 &  0 & 0 &  0 & 0 \cr
\theta^{(1)}_1 & 0 & 0 &  0 & 0 &  1 & 1 &  1 & 1 \cr
\theta^{(2)}_0 & 2 & 1 &  1 & 0 &  2 & 1 &  1 & 0 \cr
\theta^{(2)}_1 & 0 & 1 &  1 & 2 &  0 & 1 &  1 & 2 \cr}.
$$
The columns of this matrix represent the monomials
in the parametrization (\ref{parametrization}). The model
$\mathcal{M}$ lies in the $5$-dimensional subsimplex of $\Delta_7$
given by $\, p_{001} = p_{010}\,$ and $\,p_{101} = p_{110} $,
and it consists of all rank one matrices
$$ \begin{pmatrix}
      p_{000} & p_{001} &  p_{100} & p_{101} \\
      p_{010} & p_{011} & p_{110} & p_{111} \\
 \end{pmatrix}.
 $$
In algebraic geometry, the surface $\mathcal{M}$ is called a {\em rational normal scroll}.
 \qed
\end{example}

The matrix $A$ has repeated columns whenever $s_i \geq 2$ for some
$i$. It is sometimes convenient to represent the model
$\mathcal{M}$ by the matrix $\tilde{A}$ which is obtained from $A$
by removing repeated columns.  We label the columns of the matrix
$\tilde{A}$ by elements $v  = (v^{(1)}, \ldots, v^{(k)}) $ of
(\ref{statespace}) whose components $v^{(i)} \in
\{0,1,\ldots,t_i\}^{s_i}$ are weakly increasing. Hence $\tilde{A}$
is a $d \times {\tilde n}$-matrix with
\begin{equation}
\label{nprime} {\tilde n} \quad = \quad  \prod_{i=1}^k \binom{s_i + t_i}{s_i}.
\end{equation}
The model $\mathcal{M}$ and its mixtures are subsets of a subsimplex
$\Delta_{{\tilde n}-1}$ of $\Delta_{n-1}$.

We now introduce {\em marginal likelihood integrals}. All our
domains of integration in this paper are polytopes that
are products of standard probability simplices. On each
such polytope we fix the standard Lebesgue probability measure.
In other words, our discussion of Bayesian inference refers to the
uniform prior on each parameter space. Naturally, other prior distributions,
such as Dirichlet priors, are of interest, and our methods
are extended to these in Section 5. In what follows, we
simply work with uniform priors.

We identify the state space (\ref{statespace}) with the set
$\{1,\ldots,n\}$. A {\em data vector} $U=(U_1,\ldots,U_n)$ is thus
an element of $\N^n$. The {\em sample size} of these data is $U_1
+ U_2 + \cdots + U_n = N$. If the sample size $N$ is fixed then
the probability of observing these data is
\begin{equation*}
{\bf L}_U(\theta) \quad = \quad
\frac{N !}{U_1 ! U_2 ! \cdots U_n !} \cdot
 p_1(\theta)^{U_1} \cdot
 p_2(\theta)^{U_2} \cdot
 \cdots  \cdot
 p_n(\theta)^{U_n}.
 \end{equation*}
This expression is a function on the polytope $P$
which is known as the {\em likelihood function} of the
data $U$ with respect to the independence model $\mathcal{M}$.
The {\em marginal likelihood} of the data $U$ with respect to the
model $\mathcal{M}$ equals
\begin{equation*}
 \int_P {\bf L}_U(\theta) \, d \theta .
 \end{equation*}
The value of this integral is a rational number which we
now compute explicitly. The data $U$ will enter this calculation by way of
the {\em sufficient statistic} $\,b= A \cdot U $, which is a
vector in $ \N^d $. The coordinates of this vector are denoted
$b^{(i)}_j$ for $i=1,\ldots,k$ and $j=0,\ldots,t_k$. Thus
$b^{(i)}_j$ is the total number of times the value
$j$ is attained by one of the random variables
$X^{(i)}_1,\ldots,X^{(i)}_{s_i}$ in the $i$-th group. Clearly, the
sufficient statistics satisfy
\begin{equation}
\label{vectorcomponent} b^{(i)}_0 +  b^{(i)}_1 + \cdots +
b^{(i)}_{t_i} \,\,\, = \,\,\, s_i \cdot N \qquad \hbox{for
all}\,\,\, i = 1,2,\ldots,k .
\end{equation}
The likelihood function  ${\bf L}_U(\theta)$ is the constant
$\,\frac{N !}{U_1 ! \cdots U_n !} \,$ times the monomial
 $$  \theta^b \,\,\, = \,\,\, \prod_{i=1}^k \prod_{j=0}^{t_i}  (\theta^{(i)}_j)^{b^{(i)}_j}. $$
 The logarithm of this function   is concave on the polytope $P$,
 and its maximum value is attained at the point $\hat \theta$
with coordinates  $ \, {\hat \theta }^{(i)}_j \, = \, b^{(i)}_j/(s_i \cdot N)$.

\begin{lemma}
\label{toricintegral} The integral of the monomial
$\theta^b$ over the polytope $P$ equals
$$ \int_P \theta^b d \theta \quad = \quad
\prod_{i=1}^k \frac{t_i! \, b^{(i)}_0 ! \, b^{(i)}_1 ! \, \cdots \,
b^{(i)}_{t_i} !} {(s_i N + t_i) !}.$$ The product of this number
with the multinomial coefficient $\, N !/(U_1 ! \cdots U_n !)  \,$
equals the marginal likelihood of the data $U$ for the
independence model $\mathcal{M}$.
\end{lemma}

\begin{proof}
Since  $P$ is the product of simplices (\ref{prodofsimp}), this follows
from the formula
\begin{equation}
\label{gammaproduct}
\int_{\Delta_t} \!\! \theta_0^{b_0} \theta_1^{b_1} \cdots \theta_t^{b_t} d\theta
\quad = \quad
\frac{t! \cdot b_0 ! \cdot b_1 ! \, \cdots \,b_t !}{(b_0 + b_1 + \cdots + b_t + t)!}
\end{equation}
for the integral of a monomial over the standard probability simplex $\Delta_t$.
\end{proof}

Our objective is to compute marginal likelihood integrals for the mixture model
$\mathcal{M}^{(2)}$.  The natural parameter space of this model is the polytope
$$ \Theta \,\, = \,\, \Delta_1\, \times \, P \, \times \, P . $$
Let $a_v \in \N^d$ be the column vector of $A$
indexed by the state $v$, which is either in (\ref{statespace})
or in $\{1,2,\ldots,n\}$. The parametrization
(\ref{parametrization}) can be written simply as  $\,p_v = \theta^{a_v}$.
The mixture model $\mathcal{M}^{(2)}$ is defined to be the subset of $\Delta_{n-1}$
with the parametric representation
\begin{equation}
\label{mixturepara}
 p_v \,\,\, = \,\,\, \sigma_0 \cdot \theta^{a_v} \,+ \,\sigma_1 \cdot \rho^{a_v}
\qquad \hbox{for $\,(\sigma, \theta, \rho) \,\in\, \Theta $. }
\end{equation}
The likelihood function of a data vector $U \in \N^n$ for
the model $\mathcal{M}^{(2)}$ equals
\begin{equation}
\label{likelihoodfunction2}
{\bf L}_U(\sigma,\theta,\rho) \quad = \quad
\frac{N !}{U_1 ! U_2 ! \cdots U_n !} \,
 p_1(\sigma,\theta,\rho)^{U_1} \,  \cdots  \,
 p_n(\sigma,\theta,\rho)^{U_n}.
 \end{equation}
 The {\em marginal likelihood} of the data $U$ with respect to the
model $\mathcal{M}^{(2)}$ equals
\begin{equation}
\label{marginallikelihood2}
 \int_\Theta {\bf L}_U(\sigma,\theta,\rho) \, d \sigma d \theta d \rho \,\,\, = \,\,\,
\frac{N !}{U_1! \cdots U_n !} \,
\int_\Theta \prod_v (\sigma_0 \theta^{a_v} + \sigma_1 \rho^{a_v})^{U_v}
d \sigma \, d\theta \,d \rho.
\end{equation}
The following proposition shows that we can evaluate this integral {\em exactly}.

\begin{prop}\label{pr:rational}
The marginal likelihood (\ref{marginallikelihood2}) is a rational number.
\end{prop}

\begin{proof}
The likelihood function ${\bf L}_U$ is a
$\Q_{\geq 0}$-linear combination of monomials $\,\sigma^a \theta^b \rho^c $.
The integral (\ref{marginallikelihood2})
is the same $\Q_{\geq 0}$-linear combination of the numbers
\begin{equation*}
 \int_\Theta \sigma^a \theta^b \rho^c d \sigma \, d\theta \, d \rho
\quad = \quad
\bigl(\int_{\Delta_1} \! \!\sigma^a d \sigma \bigr) \cdot
\bigl(\int_P \! \theta^b d \theta \bigr) \cdot
\bigl(\int_P \! \rho^c d \rho \bigr) .
\end{equation*}
Each of the three factors is an easy-to-evaluate rational number,
by  (\ref{gammaproduct}).
\end{proof}


\begin{example}
The integral (\ref{swissintegral}) expresses the
marginal likelihood of a $4 \times 4$-table of counts
$U = (U_{ij})$ with respect to the mixture model $\mathcal{M}^{(2)}$.
Specifically, the marginal likelihood of the data
(\ref{swisstable}) equals the normalizing constant
$\,40 ! \cdot (2!)^{-12} \cdot (4!)^{-4}\,$ times the number (\ref{swissnumber}).
The model  $\mathcal{M}^{(2)}$
 consists of all non-negative $4 \times 4$-matrices
of rank $\leq 2$ whose entries sum to one. Here the parametrization
(\ref{mixturepara}) is not identifiable
because ${\rm dim}(\mathcal{M}^{(2)} ) = 11$
but ${\rm dim}(\Theta) = 13 $.
In this example,
$k = 2$, $s_1 {=} s_2 {=} 1$, $t_1 {=} t_2 {=} 3$, $d = 8$,
$n = 16$. \qed
\end{example}

In algebraic geometry, the model $\mathcal{M}^{(2)}$ is known as
the first secant variety of the Segre-Veronese variety
(\ref{segreveronese}). We could also consider the higher secant
varieties $\mathcal{M}^{(l)}$, which correspond to mixtures of
$l$ independent distributions, and much of our analysis can be
extended to that case, but for simplicity we restrict ourselves to
$l = 2$. The variety $\mathcal{M}^{(2)}$ is embedded in the
projective space $\PP^{{\tilde n}-1}$ with ${\tilde n}$ as in
(\ref{nprime}). Note that ${\tilde n}$ can be much smaller than $n
$.  If this is the case then it is convenient to aggregate states
whose probabilities are identical and to represent the data by a
vector $\tilde U \in \N^{{\tilde n}}$. Here is an example.

\begin{example}
\label{cointoss}
Let $k{=}1$, $s_1 {=} 4$ and $t_1{=}1$, so $\mathcal{M}$ is
the independence model for four identically distributed
binary random variables. Then $d = 2$ and $n = 16$. The
corresponding integer matrix and its row and column labels are
$$ A \,\, = \,\,
\bordermatrix{ & p_{0000} & p_{0001} \! & \! p_{0010} \! & \! p_{0100} \! & \! p_{1000} \!
& p_{0011} & \cdots & p_{1110} & p_{1111} \cr
\theta_0 & 4 & 3 & 3 & 3 & 3 & 2 & \cdots & 1 & 0 \cr
\theta_1 & 0 & 1 & 1 & 1 & 1 & 2 & \cdots & 3 & 4 \cr
}.
$$
However, this matrix has only ${\tilde n}=5$ distinct columns, and we instead use
$$ \tilde{A} \,\,\, = \,\,\,
\bordermatrix{ & p_0 & p_1 & p_2 & p_3 & p_4 \cr
\theta_0 & 4 & 3 & 2 & 1 & 0 \cr
\theta_1 & 0 & 1 & 2 & 3 & 4 \cr
}.
$$
The mixture model $\mathcal{M}^{(2)}$ is the subset of $\Delta_4$
given by the parametrization
$$
p_i \quad = \quad \binom{4}{i} \cdot \bigl(
 \sigma_0 \cdot \theta_0^{4-i} \cdot \theta_1^i \,\, + \,\,
 \sigma_1 \cdot \rho_0^{4-i} \cdot \rho_1^i \bigr)
 \qquad \hbox{for $i =0,1,2,3,4$.} $$
In algebraic geometry, this threefold is the secant variety of
the rational normal curve in $\PP^4$. This is the cubic hypersurface
with the implicit equation
$$ {\rm det} \bmatrix
12 p_0 & 3  p_1 & 2 p_2 \\
 3 p_1 & 2 p_2 & 3 p_3 \\
 2 p_2 & 3 p_3 & 12 p_4
\endbmatrix \quad = \quad 0.
$$
In \cite[Example 9]{HKS} the likelihood function  (\ref{likelihoodfunction2}) was studied for the data vector
$$ \tilde U \quad = \quad (\tilde U_0, \tilde U_1, \tilde U_2, \tilde U_3, \tilde U_4) \quad = \quad (51, 18, 73, 25, 75). $$
It has three local maxima (modulo swapping $\theta$ and $\rho$) whose coordinates
are algebraic numbers of degree $12$.
Using the methods to be described in the
next two sections, we computed the exact
value of the marginal likelihood for the data $\tilde U$ with respect to $\mathcal{M}^{(2)}$.
The rational number (\ref{marginallikelihood2})  is found to be the ratio of two relatively prime
integers having  $530$ digits and $552$ digits, and its numerical
value is approximately   $\,    0.7788716338838678611335742
 \cdot 10^{-22} $. \qed
\end{example}

\section{Summation over a Zonotope}

Our starting point is the observation that the Newton polytope of the
likelihood function (\ref{likelihoodfunction2}) is a zonotope.
Recall that the {\em Newton polytope} of a polynomial is the
convex hull of all exponent vectors appearing in the
expansion of that polynomial, and a polytope is a
{\em zonotope} if it is the image of a standard cube under a linear map. See \cite[\S 7]{CLO} and \cite[\S 7]{Z} for further discussions.
We are here considering the zonotope
$$
Z_A(U) \,\,\,= \,\,\, \sum_{v=1}^n U_v \cdot  [0,a_v],
$$
where $[0,a_v]$ represents the line segment between the origin and the point $a_v \in \R^d$,
 and the sum is a Minkowski sum of
line segments.  We write $Z_A = Z_A(1,1,\ldots,1)$ for the basic zonotope
which is spanned by the vectors $a_v$. Hence $Z_A(U)$ is obtained by stretching $Z_A$
along those vectors by factors $U_v$ respectively.
 Assuming that the counts $U_v$ are all positive, we have
\begin{equation}
\label{zonotopedim}
{\rm dim}(Z_A(U)) \,\,\, = \,\,\,
{\rm dim}(Z_A) \,\,\, = \,\,\, {\rm rank}(A) \,\,\, = \,\,\, d-k+1.
\end{equation}
The zonotope $Z_A$ is related to the polytope $P = {\rm conv}(A) $ in (\ref{prodofsimp}) as follows.
The dimension $d-k = t_1+ \cdots + t_k$ of $P$ is one less than ${\rm dim}(Z_A)$, and $P$ appears
as the {\em vertex figure} of the zonotope $Z_A$ at the distinguished vertex~$0$.

\begin{remark}
For higher mixtures $\mathcal{M}^{(l)}$, the Newton polytope of the likelihood function  is isomorphic to the Minkowski sum of $(l-1)$-dimensional simplices
in $\R^{(l-1)d}$. Only when $l=2$, this Minkowski sum is a zonotope.
\qed
\end{remark}

The marginal likelihood (\ref{marginallikelihood2})
we wish to compute is the integral
\begin{eqnarray}
\label{sec3integral}
\int_{\Theta}\prod_{v=1}^n (\sigma_0 \theta^{a_v} + \sigma_1
\rho^{a_v})^{U_v}d\sigma d\theta d\rho
\end{eqnarray}
times the constant $N !/ (U_1 ! \cdots U_n !)$. Our approach to this computation is to sum over the lattice points in the zonotope $Z_A(U)$. If the matrix $A$ has repeated columns, we may replace $A$ with the reduced matrix ${\tilde A}$ and $U$ with the corresponding reduced data vector
${\tilde U}$. If one desires the marginal likelihood for the reduced data vector ${\tilde U}$ instead of  the original data vector $U$, the integral remains the same while the normalizing constant becomes
$$
\frac{N!}{{\tilde U}_1 ! \cdots {\tilde U}_{\tilde n} !} \cdot \alpha_1^{{\tilde U}_1}
\cdots \alpha_{\tilde n}^{{\tilde U}_{\tilde n}},
$$
where $\alpha_i$ is the number of columns in $A$ equal to the $i$-th column of ${\tilde A}$.
In what follows we ignore the normalizing constant and focus on computing the
integral (\ref{sec3integral}) with respect to the original matrix $A$.

For  a vector $b \in \mathbb{R}^d_{\geq 0}$ we let
$|b|$ denote its $L^1$-norm $\sum_{t=1}^d b_t$. Recall from (\ref{parametrization}) that all
 columns of the  $d \times n$-matrix $A$ have the same coordinate sum
   $$
a \,\,\, := \,\,\, |a_v|\,\,\,=\,\,\, s_1 + s_2 + \cdots + s_k, \,\,\,\mbox{ for all } v = 1,2,\ldots,n,
$$
and from (\ref{vectorcomponent}) that we may denote the entries of
a vector $b \in \R^d$ by $b_j^{(i)}$ for $i = 1, \ldots, k$ and $j
= 0,\ldots,t_k$. Also, let $\mathbb{L}$ denote the image of the
linear map $A : \Z^n \rightarrow \Z^d$. Thus $\mathbb{L}$ is a
sublattice of rank $d-k+1$ in $\Z^d$. We abbreviate $\,
Z^{\mathbb{L}}_A(U) \,:=\, Z_A(U) \cap \mathbb{L}$.  
Now, using the binomial theorem, we have
$$
(\sigma_0 \theta^{a_v} + \sigma_1
\rho^{a_v})^{U_v}\,\,=\,\,\sum_{x_v=0}^{U_v}\binom{U_v}{x_v}\sigma_0^{x_v}\sigma_1^{U_v-x_v}\theta^{x_v\cdot
a_v }\rho^{(U_v-x_v)\cdot a_v}.
$$
Therefore, in the expansion of the integrand in (\ref{sec3integral}),
the exponents of 
$\theta$ are of the form of $b = \sum_v x_v a_v\in
Z_A^\LL(U) ,\,\, 0 \leq x_v\leq U_v $. The other exponents may be expressed in terms of $b$. This gives us 
\begin{equation}
\label{integrandexpansion} \prod_{v=1}^{n} (\sigma_0 \theta^{a_v} +
\sigma_1 \rho^{a_v})^{U_v} \,\,= \sum_{\substack{b \in
Z^{\mathbb{L}}_A(U) \\ c=AU-b}} \!\! \! \phi_A(b,U) \cdot
\sigma_0^{|b|/a}\cdot \sigma_1^{|c|/a} \cdot \theta^b \cdot
\rho^{c}.
\end{equation}
Writing $\,
 {\bf D}(U) \, = \,\{(x_1,\ldots,x_n)\in \Z^n\,:\,
0\leq x_v\leq U_v, v=1,\ldots,n\} $,
 we can see that the coefficient in (\ref{integrandexpansion}) equals
\begin{equation}\label{sumproduct}
\phi_A(b,U) \,\,\, = \,\, \sum_{\substack{Ax=b \\  x \in {\bf
D}(U) }} \prod_{v=1}^n \binom{U_v}{x_v}.
\end{equation}
Thus, by formulas (\ref{gammaproduct}) and (\ref{integrandexpansion}), the integral (\ref{sec3integral})
evaluates to
\begin{eqnarray}\label{eq:sum}
\sum_{\substack{b \in Z^{\mathbb{L}}_A(U) \\ c = AU-b}} \!\!
\phi_A(b,U)\cdot\frac{(|b|/a)!\,(|c|/a)!}{(|U|+1)!}\cdot\!
\prod_{i=1}^{k}\left( \frac{t_i!\,b_0^{(i)}!\,\cdots\,
b_{t_i}^{(i)}!}{(|b^{(i)}|+t_i)!}\,\,
\frac{t_i!\,c_0^{(i)}!\,\cdots\,
c_{t_i}^{(i)}!}{(|c^{(i)}|+t_i)!}\right)\!.
\end{eqnarray}
We summarize the result of this derivation in the following theorem.

\begin{thm} \label{thm:formula}
The marginal likelihood of the data $U$ in the mixture model
$\mathcal{M}^{(2)}\!$ is equal to the sum (\ref{eq:sum})
times the normalizing constant $N !/ (U_1 ! \cdots U_n !)$.
\end{thm}

Each individual summand in the formula (\ref{eq:sum}) is a ratio of factorials
and hence can be evaluated symbolically. The challenge in turning
Theorem \ref{thm:formula} into a practical algorithm lies in the fact
that both of the sums (\ref{sumproduct}) and (\ref{eq:sum})
are over very large sets. We shall discuss these challenges
and present techniques from both computer science and
mathematics for addressing them.

We first turn our attention to the coefficients $\phi_A(b,U)$ of
the expansion (\ref{integrandexpansion}). These quantities are
written as an explicit sum in (\ref{sumproduct}). The first useful
observation is that these coefficients are also the coefficients
of the expansion
\begin{equation}
\label{expansion} \prod_{v} (\theta^{a_v} + 1)^{U_v} \quad =
\,\,\, \sum_{b \in Z^{\mathbb{L}}_A(U)} \!\! \phi_A(b,U) \cdot
\theta^b,
\end{equation}
which comes from substituting $\sigma_i = 1$ and $\rho_j = 1$ in
(\ref{integrandexpansion}). 
When the cardinality of $Z^{\mathbb{L}}_A(U)$ is sufficiently small, the
quantity $\phi_A(b,U)$ can be computed quickly by
expanding (\ref{expansion}) using a computer algebra system. We
used {\tt Maple} for this purpose and all other symbolic
computations in this project.

If the expansion (\ref{expansion}) is not feasible, then it is
tempting to compute the individual $\phi_A(b,U)$ via the
sum-product formula (\ref{sumproduct}). This method requires
summation over the set $\, \{x \in {\bf D}(U) \, : \, A x = b \}$,
which is the set of lattice points in an $(n-d+k-1)$-dimensional
polytope. Even if this loop can be implemented, performing the sum
in (\ref{sumproduct}) symbolically requires the evaluation of many
large binomials, which causes the process to be rather
inefficient.

An alternative is offered by the following recurrence formula:
\begin{equation}\label{eq:recurrence}
\phi_A(b,U)\,\,\, = \,\,\,
\sum_{x_n=0}^{U_n}\binom{U_n}{x_n}\phi_{A\setminus a_n}(b-x_n a_n
, U\setminus U_n).
\end{equation}
 This is equivalent to writing the integrand in
(\ref{sec3integral})~as
$$
\left(\prod_{v=1}^{n-1} (\sigma_0 \theta^{a_v} + \sigma_1
\rho^{a_v})^{U_v}\right) (\sigma_0 \theta^{a_n} + \sigma_1
\rho^{a_n})^{U_n}.
$$
More generally, for each $0<i<n$, we have the recurrence
\begin{equation*}
\phi_A(b,U) = \sum_{b' \in
Z^{\mathbb{L}}_{A'}(U')} \!\! \phi_{A'}(b', U') \cdot
\phi_{A\setminus A'}(b - b', U\setminus U'),
\end{equation*}
where  $A'$ and $U'$ consist of the first $i$ columns and entries of
$A$ and $U$ respectively. This corresponds to the factorization
$$
\left(\prod_{v=1}^{i} (\sigma_0 \theta^{a_v} + \sigma_1
\rho^{a_v})^{U_v}\right) \left(\prod_{v=i+1}^{n} (\sigma_0 \theta^{a_v} + \sigma_1
\rho^{a_v})^{U_v}\right).
$$
This formula gives flexibility in designing algorithms with different payoffs in time and space complexity,
to be discussed in Section 4.

The next result records useful facts about the quantities $\phi_A(b,U)$.

\begin{prop}\label{pr:3} Suppose $b \in \Z^{\LL}_A(U)$ and $c = AU-b$. Then, the following quantities are all equal to $\phi_A(b,U)$: \\
(1) $\# \bigl\{z \in \{0,1\}^{N} : A^U z=b \bigr\},$
 where $A^U$ is the extended matrix
$$ A^{U} \,\, := \,\, (
\underbrace{a_1,\ldots,a_1}_{U_1},\underbrace{a_2,\ldots,a_2}_{U_2},\ldots,\underbrace{a_n,\ldots,a_n}_{U_n}),
$$
(2) $\phi_A(c,U)$, \\
(3) \begin{equation*}
\sum_{\substack{Ax=b \\l_j \leq x_j\leq
u_j }} \prod_{v=1}^n \binom{U_v}{x_v},
\end{equation*}
where $\,u_j=\min\,\,\{U_j\} \cup \{b_m/a_{jm}\}_{m=1}^{n}\,$ and $\,l_j=U_j - \min\,\,\{U_j\} \cup \{c_m/a_{jm}\}_{m=1}^{n} \,$.
\end{prop}

\begin{proof}
\noindent (1) This follows directly from (\ref{expansion}).

\noindent (2) For each $z \in \{0,1\}^{N}$ satisfying $A^Uz=b$, note that $\bar z = (1, 1, \ldots, 1) - z$ satisfies $A^U\bar z = c$, and vice versa. The conclusion thus follows from (1).

\noindent (3)  We require $Ax=b$ and $x\in {\bf D}(U)$. If
$x_j>u_j= b_m/a_{jm}$ then $a_{jm}x_j>b_m$, which implies
$Ax\neq b$.  The lower bound is derived by a similar argument.
\end{proof}

One aspect of our approach is the decision, for any given
model $A$ and data set $U$, whether or not to attempt the
expansion (\ref{expansion}) using computer algebra. This decision
depends on the cardinality of the set $\,Z^{\mathbb{L}}_{A}(U)$.
In what follows, we compute the number exactly when $A$ is
unimodular. When $A$ is not unimodular, we obtain useful
lower and upper bounds for $\# Z^{\mathbb{L}}_{A}(U)$.

Let $S$ be any subset of the columns of $A$. We call $S$
 {\em independent} if its elements are linearly independent in $\R^d$.
With $S$ we associate the integer
$$ {\rm index}(S) \,\,\, := \,\,\, [\R S \cap \LL : \Z S]. $$
This is the index of the abelian group generated by $S$ inside
the possibly larger abelian group of all lattice points in $\LL = \Z A$ that lie in the
span of $S$.
The following formula is due to R.~Stanley
and appears in \cite[Theorem 2.2]{St3}:

\begin{prop}
The number of lattice points in the zonotope $Z_A(U)$ equals
\begin{equation}\label{zonotopedimension}
\# Z_{A}^\LL(U)  \,\,\,=\,\,\, \sum_{S \subseteq A \,\, {\rm
indep.}} \!\!
 {\rm index}(S) \cdot \prod_{a_v\in S} U_v.
\end{equation}
\end{prop}

In fact, the number of monomials in (\ref{integrandexpansion})
equals $\#M_A(U)$, where $M_A(U)$ is the set $\{b\in Z_A^\LL(U)\, :\, \phi_A(b,U)\neq 0\}$, and
this set can be different from $Z_A^\LL(U)$. For that number
we have the following upper and lower bounds.
The proof of Theorem \ref{thm:uplow} will be omitted here.
It uses the methods in \cite[\S 2]{St3}.

\begin{thm}\label{thm:uplow}
The number $\#M_A(U)$ of monomials in the expansion
 (\ref{integrandexpansion}) of the likelihood function to be integrated satisfies
 the two inequalities
\begin{equation}
\label{eq:uplow}
\sum_{S \subseteq A \,\, {\rm indep.}}  \prod_{v\in S}
U_v\,\,\, \leq\,\,\, \#M_A(U)\,\,\,\leq\,\, \sum_{S \subseteq A \,\, {\rm indep.}} \!\!
 {\rm index}(S) \cdot \prod_{v\in S} U_v.
\end{equation}
\end{thm}

By definition, the matrix $A$ is {\em unimodular} if
${\rm index}(S) = 1$ for all independent subsets $S$ of the columns of $A$.
In this case, the upper bound coincides with the
lower bound, and so $M_A(U)=Z_{A}^\LL(U)$. This happens in the classical case of two-dimensional contingency tables ($k=2$ and $s_1 = s_2 = 1$). In general, $\#Z_{A}^\LL(U)/\#M_A(U)$ tends to $1$ when all coordinates of $U$ tend to infinity. This is why we believe that $\#Z_A^\LL(U)$ is a good approximation of $\#M_A(U)$. For computational purposes, it suffices to know $\#Z_{A}(U)$.

\begin{remark}
There exist integer matrices $A$ for which $\# M_A(U)$ does not agree
with the upper bound in Theorem \ref{thm:uplow}. However, we conjecture
that $\# M_A(U)=\# Z_A^\LL(U)$ holds for matrices $A$ of
Segre-Veronese type as in (\ref{parametrization}) and strictly positive data vectors $U$. \qed
\end{remark}

\begin{example} Consider the {\em $100$ Swiss Francs} example in Section \ref{introduction}.
Here $A$ is unimodular and it has $16145$ independent subsets $S$.
The corresponding sum of $16145$ squarefree monomials in
(\ref{zonotopedimension}) gives the number of terms in the
expansion of (\ref{swissintegrand}).
  For the data $U$ in (\ref{swisstable}) this sum evaluates to
  $   3,892,097$. \qed
\end{example}
\begin{example}
We consider the matrix and data from Example \ref{cointoss}.
\begin{eqnarray*}
\tilde A \,\,\, &=& \,\,\,
\begin{pmatrix}
      0 & 1 &  2 & 3 & 4 \\
      4 & 3 &  2 & 1 & 0 \\
\end{pmatrix} \\
\tilde U\,\,\,&=&\,\,\,
\begin{pmatrix}
51, 18, 73, 25, 75
\end{pmatrix}
\end{eqnarray*}
By Theorem \ref{thm:uplow}, the lower bound is 22,273 and the
upper bound is 48,646. Here the number $\#M_{\tilde A}(\tilde U)$
of monomials agrees with the latter. \qed
\end{example}

We next  present a formula for ${\rm index}(S)$ when $S$ is any
 linearly independent subset of the columns of the matrix $A$.
 After relabeling we may assume that
$S = \{a_1,\ldots,a_k\}$ consists of the first $k$ columns of $A$.
Let $H = VA$ denote the row Hermite normal form of $A$.
Here $V \in SL_d(\Z)$ and $H$ satisfies
$$ H_{ij} = 0 \hbox{ for } i>j \,\, \hbox{ and }\,\,
0 \leq  H_{ij} < H_{jj}  \hbox{ for } i < j.
$$
Hermite normal form is a built-in function
in computer algebra systems. For  instance,
in {\tt Maple} the command is {\tt ihermite}.
Using the invertible matrix $V$, we may replace
$A$ with $H$, so that  $\R S$ becomes $\R^k$ and
$\Z S$ is the image over $\Z$ of the upper left $k \times k$-submatrix of $H$.
We seek the index of that lattice in the possibly larger lattice $\Z A \cap \Z^k$.
To this end we compute
the column Hermite normal form $H' =  H V'  $.
Here $V' \in SL_n(\Z)$ and $H'$ satisfies
$$ H'_{ij} = 0 \hbox{ if } i>j \hbox{ or } j > d \, \,\, \hbox{ and } \,\,
0 \leq H_{ij} < H_{ii} \hbox{ for } i < j . $$
The lattice $\Z A \cap \Z^k $ is spanned by the first $k$ columns of $H'$,
and this implies
$$
{\rm index}(S) \quad = \quad \frac{H_{11} H_{22}\, \cdots\, H_{kk}}
{H'_{11} H'_{22}\, \cdots\, H'_{kk}}.
$$

\section{Algorithms}

In this section we discuss algorithms for computing the integral (\ref{sec3integral}) exactly,
and we discuss their advantages and limitations. In particular, we examine four main techniques which represent
the formulas  (\ref{eq:sum}), (\ref{expansion}), (\ref{zonotopedim}) and (\ref{eq:recurrence}) respectively.
The practical performance of the various algorithms
is compared by computing the integral in Example \ref{cointoss}.

A {\tt Maple} library which implements our algorithms is made available at
$$ \hbox{\url{http://math.berkeley.edu/~shaowei/integrals.html}.} $$
The input for our {\tt Maple} code consists of parameter vectors
$s = (s_1, \ldots, s_k)$ and $t = (t_1, \ldots, t_k)$ as
well as a data vector $U \in \N^n$. This input uniquely
specifies the $d \times n$-matrix $A$.
Here $d$ and $n$ are as in (\ref{DEFdn}). The output features the matrices $A$ and $\tilde A$, the marginal likelihood integrals for $\mathcal{M}$ and $\mathcal{M}^{(2)}$, as well as the bounds in (\ref{eq:uplow}).

We tacitly assume that $A$ has been replaced with the reduced matrix $\tilde A$. Thus from now on we assume that $A$ has no repeated columns. This requires some care concerning the normalizing constants. All columns of the matrix $A$ have the same coordinate sum $a$,
and the convex hull of the columns is the polytope $\,P =
\Delta_{t_1} \times \Delta_{t_2} \times \cdots \times \Delta_{t_k}$.
Our domain of integration is the following polytope
of dimension $2d-2k+1$:
\begin{equation*}
\Theta \,\,= \,\, \Delta_1 \times P \times P.
 \end{equation*}
 We seek to compute the rational number
\begin{equation}
\label{sec4integral}
\int_{\Theta}\,\prod_{v=1}^n (\sigma_0 \theta^{a_v} + \sigma_1
\rho^{a_v})^{U_v}d\sigma d\theta d\rho,
\end{equation}
where integration is with respect to Lebesgue probability measure.
Our {\tt Maple} code outputs this integral multiplied
with the statistically correct normalizing constant.
That constant will be ignored in what follows.
In our complexity analysis, we fix $A$ while allowing the data $U$ to vary. The complexities will
 be given in terms of the sample size $N=U_1+ \cdots+U_n$.

\subsection{Ignorance is Costly}

Given an integration problem such as (\ref{sec4integral}), a first attempt is to use the symbolic integration
capabilities of a computer algebra package such as {\tt Maple}. We will refer to this method as {\it ignorant integration}:
\begin{verbatim}
U  := [51, 18, 73, 25, 75]:
f  := (s*t^4        +(1-s)*p^4        )^U[1] *
      (s*t^3*(1-t)  +(1-s)*p^3*(1-p)  )^U[2] *
      (s*t^2*(1-t)^2+(1-s)*p^2*(1-p)^2)^U[3] *
      (s*t  *(1-t)^3+(1-s)*p  *(1-p)^3)^U[4] *
      (s    *(1-t)^4+(1-s)    *(1-p)^4)^U[5]:
II := int(int(int(f,p=0..1),t=0..1),s=0..1);
\end{verbatim}

In the case of mixture models, recognizing the integral as the sum of integrals of monomials over a polytope
allows us to avoid the expensive integration step above by using (\ref{eq:sum}). To demonstrate
 the power of using (\ref{eq:sum}), we implemented a simple algorithm that computes each $\phi_A(b,U)$
 using the naive expansion in (\ref{sumproduct}). We computed the integral in Example \ref{cointoss} with a small
  data vector $U = (2,2,2,2,2)$, which is the rational number
\begin{equation*}
\frac{66364720654753}{59057383987217015339940000},
\end{equation*}
and summarize the run-times and memory usages of the two algorithms in the table below.
All experiments reported in this section are done in {\tt Maple}.
\begin{center}
\begin{tabular} {r c c}
\hline
 & Time(seconds) & Memory(bytes) \\
\hline
Ignorant Integration & 16.331 &  155,947,120 \\
\hline
Naive Expansion & 0.007 & 458,668 \\
\hline
\end{tabular}
\end{center}
 For the remaining comparisons in this section, we no longer consider the ignorant integration
 algorithm because it is computationally too expensive.

\subsection{Symbolic Expansion of the Integrand}

While ignorant use of a computer algebra system is unsuitable
for computing our integrals, we can still exploit its powerful
polynomial expansion capabilities to find the coefficients of (\ref{expansion}).
A major advantage is that it is very easy to write code for this method. We compare
the performance of this symbolic expansion algorithm against that of the naive expansion
algorithm. The table below concerns computing the coefficients $\phi_A(b,U)$ for the original data $U = (51, 18, 73, 25, 75)$. The column ``Extract'' refers to the time taken to extract the coefficients $\phi_A(b,U)$ from the expansion of the polynomial, while the column  ``Sum'' shows the time taken to evaluate (\ref{eq:sum}) after all the needed values of $\phi_A(b,U)$ had been computed and extracted.
\begin{center}
\begin{tabular} { r c c c c c }
\hline
 & \multicolumn{4}{c}{Time(seconds)}         & Memory\\
\cline{2-5}
 & $\phi_A(b,U)$ & Extract & Sum & Total & (bytes) \\
\hline
Naive Expansion &  2764.35 & - & 31.19 &  2795.54 &  10,287,268 \\
\hline
Symbolic Expansion & 28.73 &  962.86 & 29.44 & 1021.03 & 66,965,528 \\
\hline
\end{tabular}
\end{center}

\subsection{Storage and Evaluation of $\phi_A(b,U)$}
\label{recursivesection}

Symbolic expansion is 
fast for computing $\phi_A(b,U)$, but it has two  
drawbacks: high memory consumption and the long time it takes to extract the values of $\phi_A(b,U)$. One solution is to create specialized data structures and algorithms for expanding (\ref{expansion}), rather using than those offered by {\tt Maple}.

First, we tackle the problem of storing the coefficients $\phi_A(b,U)$ for $b \in Z^{\LL}_A(U) \subset \R^d$ as they are being computed. One naive method is to use a $d$-dimensional array $\phi[\cdot]$. However, noting that $A$ is not row rank full, we can use a $d_0$-dimensional array to store $\phi_A(b,U)$, where $d_0 = {\rm rank}(A) = d-k+1$. Furthermore, by Proposition \ref{pr:3}(2), the expanded integrand is a symmetric polynomial, so only half the coefficients need to be stored. We will leave out the implementation details so as not to complicate our discussions. In our algorithms, we will assume that the coefficients are stored in a $d_0$-dimensional array $\phi[\cdot]$, and the entry that represents $\phi_A(b,U)$ will be referred to as $\phi[b]$.

Next, we discuss how $\phi_A(b,U)$ can be computed. One could use the naive expansion (\ref{sumproduct}),
 but this involves evaluating many binomials coefficients and products, so the algorithm is inefficient
 for data vectors with large coordinates. A much more efficient solution uses the recurrence formula (\ref{eq:recurrence}):

\begin{algorithm}[RECURRENCE($A$, $U$)]\label{al:41} $ $ \\
{\bf Input:} The matrix $A$ and the vector $U$. \\
{\bf Output:} The coefficients $\phi_A(b,U)$. \\
{\bf Step 1}: Create a $d_0$-dimensional array $\phi$ of zeros. \\
{\bf Step 2}: For each $x \in \{0, 1,\ldots, U_1\}$ set
$$
\phi[xa_1]\,\,\,:=\,\,\, \binom{U_1}{x}.
$$
{\bf Step 3}: Create a new $d_0$-dimensional array $\phi'$. \\
{\bf Step 4}: For each $2 \leq j \leq n$ do

\hspace{1cm} 1. Set all the entries of $\phi'$ to $0$.

\hspace{1cm}  2. For each $x \in \{0, 1, \ldots, U_j\}$ do

\hspace{1.5cm}   For each non-zero entry $\phi[b]$ in $\phi$ do

\hspace{2cm}
Increment $\phi'[b+xa_j]$ by $\binom{U_j}{x}\phi[b].$

\hspace{1cm}  3. Replace $\phi$ with $\phi'$. \\
{\bf Step 5}: Output the array $\phi$.
\end{algorithm}
The space complexity of this algorithm is $O(N^{d_0})$
and its time complexity is $O(N^{d_0+1})$. By comparison,
the naive expansion algorithm has space complexity $O(N^d)$ and time complexity $O(N^{n+1})$.

We now turn our attention to computing the integral (\ref{sec4integral}). One major issue is the lack of
 memory to store all the terms of the expansion of the integrand. We overcome this problem by writing the
 integrand as a product of smaller factors which can be expanded
separately. In particular, we partition the columns of $A$ into
submatrices $A^{[1]}, \ldots, A^{[m]}$ and let $U^{[1]}, \ldots,
U^{[m]}$ be the corresponding partition of $U$. Thus the integrand
becomes
$$
\prod_{j=1}^m\prod_{v} (\sigma_0 \theta^{a^{[j]}_v} + \sigma_1
\rho^{a^{[j]}_v})^{U^{[j]}_v} ,
$$
where $a^{[j]}_v$ is the $v$th column in the matrix $A^{[j]}$.
The resulting algorithm for evaluating the integral is as follows:

\begin{algorithm}[Fast Integral] \label{fastt} $ $\\
{\bf Input:} The matrices $A^{[1]}, \ldots, A^{[m]}$,
vectors $U^{[1]}, \ldots, U^{[m]}$ and the vector $t$. \\
{\bf Output:} The value of the integral (\ref{sec4integral})
in exact rational arithmetic. \\
{\bf Step 1}: For  $1\leq j\leq m$, compute $\phi^{[j]}:={\rm
RECURRENCE}(A^{[j]},U^{[j]})$. \\
{\bf Step 2}: Set $I := 0$. \\
{\bf Step 3}: For each non-zero entry $\phi^{[1]}[b^{[1]}]$ in $\phi^{[1]}$ do

$\hspace{2cm}\vdots $ \\
\phantom{{\bf Step 4}:} For each non-zero entry $\phi^{[m]}[b^{[m]}]$ in $\phi^{[m]}$ do

\hspace{1.5cm} Set $b:= b^{[1]} + \cdots + b^{[m]}$, $c := AU-b$, $\phi:= \prod_{j=1}^{m}\phi^{[j]}[b^{[j]}]$.

\hspace{1.5cm} Increment $I$ by

\hspace{2cm} $\phi\,\cdot \,\frac{(|b|/a)!(|c|/a)!}{(|U|+1)!}\,\cdot\,
\prod_{i=1}^{k} \frac{t_i!\,b_0^{(i)}!\cdots
b_{t_i}^{(i)}!}{(|b^{(i)}|+t_i)!}\,\, \frac{t_i!\,c_0^{(i)}!\cdots
c_{t_i}^{(i)}!}{(|c^{(i)}|+t_i)!}.
$ \vspace{.2cm}
\\
{\bf Step 4}: Output the sum $I$.
\end{algorithm}
The algorithm can be sped up by precomputing the factorials used in the product in Step 3. The space and time complexity of this algorithm is $O(N^S)$ and $O(N^T)$ respectively, where
$S=\max_i {\rm rank}\, A^{[i]}$ and $T =\sum_i{\rm rank}A^{[i]}$. From this, we see that the splitting of the
integrand should be chosen wisely to achieve a good pay-off between the two complexities.

In the table below, we compare the naive expansion algorithm and the fast integral algorithm for the
data $U =(51, 18, 73, 25, 75)$. We also compare the effect of splitting the integrand into two factors,
 as denoted by $m=1$ and $m=2$. For $m=1$, the fast integral algorithm takes significantly less time than
 naive expansion, and requires only about 1.5 times more memory.

\begin{center}
\begin{tabular} { r c c }
\hline
 & Time(minutes) & Memory(bytes) \\
\hline
Naive Expansion & 43.67 &  9,173,360 \\
\hline
Fast Integral (m=1) &  1.76 & 13,497,944 \\
\hline
Fast Integral (m=2) &  139.47 & 6,355,828 \\
\hline
\end{tabular}
\end{center}

\subsection{Limitations and Applications}

\begin{figure}
\centering
\caption{Comparison of computation time against sample size.}
\label{graph1}
\includegraphics{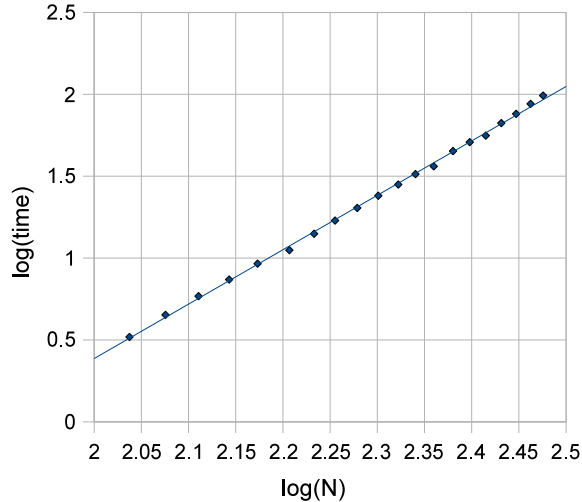}
\end{figure}

While our algorithms are optimized for exact evaluation of integrals for mixtures of independence models, they may not be practical for applications involving large sample sizes. To demonstrate their limitations, we vary the sample sizes in Example \ref{cointoss} and compare the computation times. The data vectors $U$ are generated by scaling $U=(51,18,73,25,75)$ according to the sample size $N$ and rounding off the entries. Here, $N$ is varied from $110$ to $300$ by increments of $10$. Figure \ref{graph1} shows a logarithmic plot of the results. The times taken for $N=110$ and $N=300$ are $3.3$ and $98.2$ seconds respectively. Computation times for larger samples may be extrapolated from the graph. Indeed, a sample size of $5000$ could take more than $13$ days. 

For other models, such as the {\em $100$ Swiss Francs} example in Section 1 and that of the schizophrenic patients in Example \ref{schizoex}, the limitations are even more apparent. In the table below, for each example we list the sample size, computation time, rank of the corresponding $A$-matrix and the number of terms in the expansion of the integrand. Despite having smaller sample sizes, the computations for the latter two examples take a lot more time. This may be attributed to the higher ranks of the $A$-matrices and the larger number of terms that need to be summed up in our algorithm.

\begin{center}
\begin{tabular} { r c c c c }
\hline
 & Size & Time & Rank & \#Terms \\
\hline
Coin Toss & 242 & 45 sec & 2 & 48,646 \\
\hline
$100$ Swiss Francs & 40 & 15 hrs & 7 & 3,892,097\\
\hline
Schizophrenic Patients & 132 & 16 days & 5 & 34,177,836 \\
\hline
\end{tabular}
\end{center}

Despite their high complexities, we believe our algorithms are important because they provide a gold standard with which approximation methods such as those studied in \cite{CH} can be compared. Below, we use our exact methods to ascertain the accuracy of asymptotic formula derived by Watanabe {\it et al.}~using desingularization methods from algebraic geometry \cite{W, YW1, YW2}. 

\begin{example}
Consider the model from Example \ref{cointoss}. Choose data vectors $U=(U_0,U_1,U_2,U_3,U_4)$ with $U_i = Nq_i$ where $N$ is a multiple of $16$ and 
$$
q_i =  \frac{1}{16}\binom{4}{i}, \quad i=0,1,\ldots,4.
$$
Let $I_N(U)$ be the integral (\ref{sec4integral}). Define
$$
F_N(U) =  N \sum_{i=0}^{4}q_i \log q_i-\log I_N(U).
$$
According to \cite{YW2}, for large $N$ we have the asymptotics
\begin{eqnarray} \label{coinflipasymp}
{\rm E}_U[F_N(U)] \,\,\, = \,\,\, \frac{3}{4} \log N +O(1)
\end{eqnarray}
where the expectation $E_U$ is taken over all $U$ with sample size $N$ under the distribution defined by $q = (q_0, q_1, q_2, q_3, q_4)$. Thus, we should expect
$$
F_{16+N} - F_{N} \,\, \approx \,\, \frac{3}{4} \log(16+N) - \frac{3}{4}\log N =: g(N).
$$
We compute $F_{16+N} - F_{N}$ using our exact methods and list the results below.
\begin{center}
\begin{tabular}{ccc}
\hline
$N$ & $F_{16+N}-F_N$  & $g(N)$\\
\hline
 16 & 0.21027043 & 0.225772497 \\
 32 & 0.12553837 & 0.132068444 \\
 48 & 0.08977938 & 0.093704053 \\
 64 & 0.06993586 & 0.072682510 \\
 80 & 0.05729553 & 0.059385934 \\
 96 & 0.04853292 & 0.050210092 \\
112 & 0.04209916 & 0.043493960 \\
\hline
\end{tabular}
\end{center} 
Clearly, the table supports our conclusion. 
The coefficient $3/4$ of $\log N$ in
the formula (\ref{coinflipasymp}) is known as
the {\em real log-canonical threshold} of the statistical model.
 The example suggests that our method could be
developed into a numerical technique for computing the
real log-canonical threshold.
\qed
\end{example}

\section{Back to Bayesian statistics}
In this section we discuss how the exact integration approach
presented here interfaces with issues in Bayesian statistics. The first concerns the
rather restrictive assumption that our marginal likelihood
integral be evaluated with respect to the uniform distribution
(Lesbegue measure) on the parameter space $\Theta$. It is standard
practice to compute such integrals with
respect to {\em Dirichlet priors}, and we shall now explain how
our algorithms can be extended to Dirichlet priors.
That extension is also available as a feature in our {\tt Maple} implementation.

Recall that the {\em Dirichlet distribution}  $\,{\rm Dir}(\alpha)
\,$ is a continuous probability distribution which is parametrized
by a vector $\alpha = (\alpha_0,\alpha_1,\ldots,\alpha_m)\,$ of
positive reals. It is the multivariate generalization of the beta
distribution and is conjugate prior (in the Bayesian sense) to the
multinomial distribution. This means that the probability
distribution function of  ${\rm Dir}(\alpha)$ specifies the belief
that the probability of the $i$th among $m+1$ events equals
$\theta_i$ given that it has been observed $\alpha_i-1$ times.
More precisely, the probability density function $f(\theta;
\alpha)$ of ${\rm Dir}(\alpha)$ is supported on the
$m$-dimensional simplex
$$ \Delta_m \,\,\, = \,\,\,
\bigl\{ (\theta_0,\ldots,\theta_m) \in \R_{\geq 0}^m \,:\,
\theta_0 + \cdots + \theta_m = 1 \bigr\}, $$
  and it equals
$$ f(\theta_0,\ldots,\theta_m; \alpha_0,\ldots,\alpha_m) \quad = \quad
\frac{1}{\B(\alpha)} \cdot
  \theta_0^{\alpha_0-1}  \theta_1^{\alpha_1-1}  \cdots \theta_m^
{\alpha_m-1}
  \,\,\, =: \,\,\, \frac{\theta^{\alpha-1}}{\B(\alpha)} . $$
Here the normalizing constant is the multinomial beta function
$$\B(\alpha) \quad = \quad
\frac{m!
  \Gamma(\alpha_0)
  \Gamma(\alpha_1) \cdots
  \Gamma(\alpha_m)}{\Gamma(\alpha_0 + \alpha_1 + \cdots + \alpha_m)} .
  $$
  Note that, if the $\alpha_i$ are all integers, then
  this is the rational number
  $$ \B(\alpha) \,\, = \,\,
  \frac{m!
( \alpha_0-1) ! ( \alpha_1-1) ! \cdots ( \alpha_m-1) ! }{
(\alpha_0 + \cdots + \alpha_m - 1)!}. $$ Thus the identity
(\ref{gammaproduct}) is the special case of the identity
$\,\int_{{\Delta}_m} f(\theta;\alpha) d \theta = 1\,$ for the
density of the Dirichlet distribution when all $\, \alpha_i = b_i+
1\, $ are integers.

We now return to the marginal likelihood for mixtures of
independence models. To compute this quantity with respect to
Dirichlet priors means the following. We fix positive real numbers
$ \alpha_0, \alpha_1$, and $\beta^{(i)}_j$ and $\gamma^{(i)}_j$
for $i=1,\ldots,k$ and $j=0,\ldots,t_i$. These specify Dirichlet
distributions on $\Delta_1$, $P$ and $P$. Namely, the Dirichlet
distribution on $P$ given by the $\beta^{(i)}_j$ is the product
probability measure given by taking the Dirichlet distribution
with parameters
$(\beta^{(i)}_0,\beta^{(i)}_1,\ldots,\beta^{(i)}_{t_i})$ on the $i$-th
factor $\Delta_{t_i}$ in the product (\ref{prodofsimp}) and
similarly for the $\gamma^{(i)}_j$. The resulting product
probability distribution on $\,\Theta \, = \, \Delta_1\, \times \,
P \, \times \, P \,$ is called the {\em Dirichlet distribution}
with parameters $\,(\alpha,\beta,\gamma) $. Its probability
density function is the product of the respective densities:
  \begin{equation}
  \label{productdensity}
  f(\sigma,\theta,\rho; \alpha,\beta,\gamma) \,\,\,\,= \,\,\,\,
\frac{\sigma^{\alpha-1}}{\B(\alpha)} \cdot
  \prod_{i=1}^k \frac{(\theta^{(i)})^{\beta^{(i)}-1}}{\B(\beta^
{(i)})} \cdot
  \prod_{i=1}^k \frac{(\rho^{(i)})^{\gamma^{(i)}-1}}{\B(\gamma^{(i)})} .
  \end{equation}
By the marginal likelihood with Dirichlet priors we mean the
integral
\begin{equation}
\label{marginallikelihood3}
   \int_\Theta {\bf L}_U(\sigma,\theta,\rho) \,
  f(\sigma,\theta,\rho; \alpha,\beta,\gamma)
d \sigma d \theta d \rho .
\end{equation}
This is a modification of  (\ref{marginallikelihood2}) and it
depends not just on the data $U$ and the model $\mathcal{M}^{(2)}$
but also on the  choice of Dirichlet parameters
$(\alpha,\beta,\gamma)$. When the coordinates of these parameters
are arbitrary positive reals but not integers, then the value of
the integral (\ref{marginallikelihood3}) is no longer a rational
number. Nonetheless, it can be computed exactly as follows. We
abbreviate the product of gamma functions in the denominator of
the density (\ref{productdensity}) as follows:
$$
\B(\alpha,\beta,\gamma) \,\,\, := \,\,\, \B(\alpha) \cdot
  \prod_{i=1}^k {\B(\beta^{(i)})} \cdot
  \prod_{i=1}^k {\B(\gamma^{(i)})} .
$$
Instead of the integrand (\ref{integrandexpansion}) we now need to
integrate
  $$
\sum_{\substack{b \in Z^{\mathbb{L}}_A(U) \\ c = AU-b}}  \frac{
\phi_A(b,U)}{\B(\alpha,\beta,\gamma)} \cdot \sigma_0^{|b|/a +
\alpha_0-1}\cdot \sigma_1^{|c|/a+\alpha_1-1} \cdot
\theta^{b+\beta-1} \cdot \rho^{c+\gamma-1}
$$
with respect to Lebesgue probability measure on $\Theta$. Doing
this term by term, as before, we obtain the following modification
of Theorem \ref{thm:formula}.

\begin{cor} \label{cor:formula}
The marginal likelihood of the data $U$ in the mixture model
$\mathcal{M}^{(2)}\!$ with respect to Dirichlet priors with
parameters $(\alpha,\beta,\gamma)$ equals
$$
\begin{matrix}
\frac{N !}{ U_1 ! \cdots U_n ! \cdot \B(\alpha,\beta,\gamma)} \cdot \sum_{\substack{b \in Z^
{\mathbb{L}}_A(U) \\ c = AU-b}} \phi_A(b,U)
\,\frac{\Gamma(|b|/a+\alpha_0)\Gamma(|c|/a+\alpha_1)}{\Gamma(|U|+|\alpha|)}
\qquad \qquad \qquad \qquad \qquad \qquad \\
\qquad \qquad \qquad \cdot  \prod_{i=1}^{k}
\bigl(
\frac{t_i!\Gamma(b_0^{(i)} + \beta_0^{(i)}) \cdots \Gamma(b_{t_i}^{(i)} +
\beta_{t_i}^{(i)})}{\Gamma(|b^{(i)}|+|\beta^{(i)}|)} \,\,
\frac{t_i!\Gamma(c_0^{(i)} + \gamma_0^{(i)}) \cdots \Gamma(c_{t_i}^{(i)}+
\gamma_{t_i}^{(i)})}{\Gamma(|c^{(i)}|+|\gamma^{(i)}|)}
\bigr).
\end{matrix}
$$
\end{cor}

A well-known experimental study by Chickering and Heckerman
\cite{CH} compares different methods for computing numerical
approximations of marginal likelihood integrals. The model
considered in \cite{CH} is the {\em naive-Bayes model}, which, in
the language of algebraic geometry, corresponds to arbitrary
secant varieties of Segre varieties. In this paper we considered
the first secant variety of arbitrary Segre-Veronese varieties.
In what follows we restrict our discussion to the intersection of
both classes of models, namely, to the first secant variety of
Segre varieties. For the remainder of this section we fix
$$ s_1 = s_2 = \cdots = s_k = 1 $$
but we allow $t_1,t_2,\ldots,t_k$ to be arbitrary positive
integers. Thus in the model of \cite[Equation (1)]{CH},
  we fix $r_C = 2$, and the $n$ there corresponds to our $k$.

To keep things as simple as possible, we shall fix the uniform
distribution as in Sections 1--4 above. Thus, in the notation of
\cite[\S 2]{CH}, all Dirichlet hyperparameters $\alpha_{ijk}$ are
set to $1$. This implies that, for any data $U \in \N^n$ and any
of our models, the problem of finding the maximum a posteriori (MAP)
configuration is equivalent to finding the maximum likelihood (ML)
configuration. To be precise, the {\em MAP configuration} is the
point $(\hat \sigma, \hat \theta, \hat \rho) $ in $\Theta$ which
maximizes the likelihood function $\,{\bf L}_U(\sigma,\theta,\rho)
\,$ in (\ref {likelihoodfunction2}). This maximum may not be
unique, and there will typically be many local maxima. Chickering
and Heckerman \cite[\S 3.2]{CH} use the expectation maximization
(EM) algorithm \cite[\S 1.3]{ASCB} to compute a numerical
approximation of the MAP configuration.

The Laplace approximation and the BIC score \cite[\S 3.1]{CH} are
predicated on the idea that the MAP configuration can be found
with high accuracy and that the data $U$ were actually drawn from
the corresponding distribution $p(\hat \sigma,\hat \theta, \hat
\rho)$. Let ${\bf H}(\sigma, \theta, \rho)$ denote the Hessian
matrix of the log-likelihood function $\, {\rm log}\, {\bf L}
(\sigma, \theta, \rho) $. Then the Laplace
approximation \cite[equation (15)]{CH} states that the logarithm
of the marginal likelihood can be approximated by
\begin{equation}
\label{Laplace} {\rm log}\, {\bf L} (\hat \sigma, \hat \theta,
\hat \rho) \,-\,
  \frac{1}{2} {\rm log}|{\rm det}\,{\bf H}(\hat \sigma, \hat \theta,
\hat \rho)| \,+\, \frac{2d-2k+1}{2} {\rm log}(2 \pi) .
\end{equation}
The Bayesian information criterion (BIC) suggests the coarser
approximation
\begin{equation}
\label{BIC} {\rm log}\, {\bf L} (\hat \sigma, \hat \theta, \hat
\rho) \,\,-\,\, \frac{2d-2k+1}{2} {\rm log} (N),
\end{equation}
where $N = U_1 + \cdots + U_n$ is the sample size.

In algebraic statistics, we do not content ourselves with the
output of the EM algorithm but, to the extent possible, we seek to
actually solve the likelihood equations \cite{HKS} and compute all
local maxima of the likelihood function. We consider it a
difficult problem to reliably find $(\hat \sigma, \hat \theta,
\hat \rho)$, and we are concerned about the accuracy of any
approximation like (\ref{Laplace}) or (\ref{BIC}).

\begin{example}
Consider the {\em $100$ Swiss Francs} table (\ref{swisstable})
discussed in the Introduction.
Here $k=2$, $s_1 = s_2 = 1$, $t_1 = t_2 = 3$,
the matrix $A$ is unimodular, and (\ref{segreveronese})
is the Segre embedding $\PP^3 \times \PP^3  \hookrightarrow \PP^{15}$.
The parameter space $\Theta$ is $13$-dimensional,
but the model $\mathcal{M}^{(2)}$ is $11$-dimensional,
so the given parametrization is not identifiable \cite{FHRZ}.
This means that the Hessian matrix {\bf H} is singular,
and hence the Laplace approximation (\ref{Laplace})
is not defined. \qed
\end{example}

\begin{example}
We compute (\ref{Laplace}) and (\ref{BIC}) for
the model and data in Example \ref{cointoss}.
According to \cite[Example 9]{HKS},
the likelihood function $p_0^{51} p_1^{18} p_2^{73} p_3^{25} p_4^{75}$
has three local maxima
$(\hat p_0,\hat p_1,\hat p_2,\hat p_3,\hat p_4)$ in the
model $\mathcal{M}^{(2)}$, and these translate into six local maxima
$(\hat \sigma, \hat \theta, \hat \rho)$ in the
parameter space $\Theta$, which is the $3$-cube.
The two global maxima $(\hat \sigma_0, \hat \theta_0, \hat \rho_0)$ in $\Theta$ are
\begin{eqnarray*}
  &&(0.3367691969, 0.0287713237, 0.6536073424), \\
  &&(0.6632308031, 0.6536073424, 0.0287713237).
\end{eqnarray*}
Both of these points in $\Theta$ give the same point in the model:
$$
(\hat p_0,\hat p_1,\hat p_2,\hat p_3,\hat p_4) \,\,\, = \,\,\, (0.12104, 0.25662, 0.20556, 0.10758, 0.30920).
$$
The likelihood function evaluates to $0.1395471101 \times 10^{-18}$ at this point.
The following table compares the various approximations. Here, ``Actual'' refers to the base-10 logarithm of the marginal likelihood in Example \ref{cointoss}.
\begin{center}
\begin{tabular} { r c}
\hline
BIC & -22.43100220   \\
\hline
Laplace &  -22.39666281 \\
\hline
Actual &  -22.10853411 \\
\hline
\end{tabular}
\end{center}
\vspace{-.2in}
\qed
\end{example}

The method for computing the marginal likelihood which was found
to be most accurate in the experiments of Chickering and Heckerman
is the {\em candidate method} \cite[\S 3.4]{CH}. This is a
Monte-Carlo method which involves running a Gibbs sampler. The
basic idea is that one wishes to compute a large sum, such as
(\ref{eq:sum}) by sampling among the terms rather than listing all
terms. In the candidate method one uses not the sum (\ref{eq:sum})
over the lattice points in the zonotope but the more naive sum
over all $2^N$ hidden data that would result in  the observed data
represented by $U$. The value of the sum is the number of terms,
$2^N$, times the average of the summands, each of which is easy to
compute. A comparison of the results of the candidate method with
our exact computations, as well as a more accurate version of
Gibbs sampling which is adapted for (\ref{eq:sum}), will be the
subject of a future study.

\smallskip

One of the applications of marginal likelihood integrals lies in
model selection. An important concept in that field is that of
{\em Bayes factors}. Given data and two competing models, the Bayes
factor is the ratio of the marginal likelihood integral of the
first model over the marginal likelihood integral of the second
model. In our context it makes sense to form that ratio for the
independence model $\mathcal{M}$ and its mixture
$\mathcal{M}^{(2)}$. To be precise, given any independence model,
specified by positive integers $s_1,\ldots, s_k, t_1,\ldots, t_k$
and a corresponding data vector $U \in \N^n$, the Bayes factor is
the ratio of the marginal likelihood in Lemma \ref{toricintegral}
and the marginal likelihood in Theorem  \ref{thm:formula}. Both
quantities are rational numbers and hence so is their ratio.

\begin{cor}
The Bayes factor which discriminates between the
independence model $\mathcal{M}$ and
the mixture model
$\mathcal{M}^{(2)}$ is a rational number. It can
be computed exactly using Algorithm \ref{fastt}
(and our {\tt Maple}-implementation).
\end{cor}

\begin{example}
\label{schizoex}
We conclude by applying our method to a data set taken from
the Bayesian statistics literature. Evans, Gilula and Guttman \cite[\S 3]{EGG}
analyzed the association between length of
hospital stay (in years $Y$) of $132$ schizophrenic patients
and the frequency with which they are visited by their relatives.
Their data set is the following contingency table of format $3 \times 3$:
$$ \!\!\!\! U \quad  = \quad \begin{matrix}
& 2 {\leq} Y {<} 10 & 10 {\leq} Y {<} 20 & 20 {\leq} Y & &{\it Totals} \cr
\hbox{Visited regularly \!\!} & 43 & 16 & 3 && {\it 62} \cr
\hbox{Visited rarely}      & 6 & 11 & 10  && {\it 27} \cr
\hbox{Visited never}     & 9 &  18 & 16 && {\it 43} \cr
{\it Totals}& {\it 58} & {\it 45} & {\it 29} & & {\bf 132} \cr
\end{matrix}
$$
They present estimated posterior means and variances for these data,
where {\em ``each estimate requires a $9$-dimensional integration''} \cite[p.~561]{EGG}.
Computing their integrals is essentially equivalent to ours,
for $k=2, s_1=s_2=1, t_1=t_2=2$ and $N=132$.
The authors emphasize that
{\em ``the dimensionality of the integral does present a problem''} \cite[p.~562]{EGG},
and they point out that
{\em ``all posterior moments can be calculated in closed form ....
however, even for modest $N$ these expressions are far to complicated
to be useful''} \cite[p.~559]{EGG}.

We differ on that conclusion. In our view, the
 closed form expressions in Section 3 are quite useful for modest sample size $N$.
Using Algorithm \ref{fastt}, we computed the
integral (\ref{sec4integral}). It is the rational number with numerator
$$
\begin{array}{l}
278019488531063389120643600324989329103876140805\\
285242839582092569357265886675322845874097528033\\
99493069713103633199906939405711180837568853737
\end{array}
$$
and denominator
$$
\begin{array}{l}
12288402873591935400678094796599848745442833177572204\\
50448819979286456995185542195946815073112429169997801\\
33503900169921912167352239204153786645029153951176422\\
43298328046163472261962028461650432024356339706541132\\
34375318471880274818667657423749120000000000000000.
\end{array}
$$
To obtain the marginal likelihood for the data $U$ above,
that rational number (of moderate size) still needs to be multiplied with the
normalizing constant
$$\frac{132!}{43! \cdot 16! \cdot 3! \cdot 6! \cdot 11! \cdot 10! \cdot 9! \cdot 18! \cdot 16!}.
$$
\qed
\end{example}

\bigskip \bigskip

\noindent {\bf Acknowledgements.} Shaowei Lin was supported by
graduate fellowship  from A*STAR (Agency for Science, Technology
and Research, Singapore). Bernd Sturmfels was supported by an
Alexander von Humboldt research prize and the U.S.~National
Science Foundation (DMS-0456960). Zhiqiang Xu is Supported by the
NSFC grant 10871196 and a Sofia Kovalevskaya prize awarded to Olga
Holtz.

\end{document}